# A Giant Tunneling Electroresistance Effect Driven by an Electrically Controlled Spin Valve at a Complex Oxide Interface

J. D. Burton[*] and E. Y. Tsymbal[**]

*Department of Physics and Astronomy, Nebraska Center for Materials and Nanoscience,
University of Nebraska, Lincoln, Nebraska 68588-0299, USA*

A giant tunneling electroresistance effect may be achieved in a ferroelectric tunnel junction by exploiting the magnetoelectric effect at the interface between a ferroelectric barrier and magnetic $La_{1-x}Sr_xMnO_3$ electrode. Using first-principles density functional theory we demonstrate that a few magnetic monolayers of $La_{1-x}Sr_xMnO_3$ near the interface act, in response to ferroelectric polarization reversal, as an atomic scale spin-valve by filtering spin-dependent current. This effect produces more than an order of magnitude change in conductance, and thus constitutes a giant resistive switching effect.

The control and utilization of the charge, spin and orbital degrees of freedom in complex oxides is one of the most promising avenues for the future of electronic devices.[1] This route has lately become a highly active research area because of advances in growing well-defined interfaces between complex oxide materials with atomic-scale precision, allowing for the engineering and cross-coupling of their unique magnetic, ferroelectric, and transport properties.[2,3] In particular recent experimental and theoretical studies of perovskite ferroelectric oxide thin films have demonstrated that ferroelectricity persists down to the nanometer scale, opening the possibility to utilize ferroelectric polarization as a functional parameter in nanoscale devices.[4]

One of the interesting implementations of such a device is a Ferroelectric Tunnel Junction (FTJ), consisting of two metal electrodes separated by a few-nm-thick ferroelectric barrier through which electrons can quantum-mechanically tunnel.[5] A FTJ exhibits the fundamental property of Tunneling ElectroResistance (TER): a change in resistance with reversal of ferroelectric polarization. For technological applications a large and reproducible TER effect is the decisive characteristic of a FTJ for non-destructive readout operation.[6] Based on simple models it was predicted that TER in FTJs can be sizable due to the dependence of the tunneling potential barrier on ferroelectric polarization orientation.[7,8] This mechanism has been elaborated using first-principles calculations showing, in addition, the importance of interface bonding and barrier decay rate effects on TER.[9-12] Very recently three experimental groups have independently reported observations of the TER effect associated with the switching of ferroelectric polarization of $BaTiO_3$[13-15] and $Pb_{1-x}Zr_xTiO_3$[16] ferroelectric films. These experimental results demonstrate the capability of perovskite-oxide ferroelectrics to serve as switchable tunnel barriers.

In parallel with the endeavor to explore FTJs, there have recently been separate efforts toward manipulating magnetization by electric fields.[17,18] Such magnetoelectric effects can be electronically induced at the surfaces and interfaces of ferromagnetic metals.[19-22] The incorporation of ferroelectric materials is especially helpful in this regard because their switchable electrical polarization can induce a large magnetoelectric response at the interface with a magnetic material. In this Letter we demonstrate, using first-principles density functional calculations, how such a magnetoelectric interaction [23-25] between a ferroelectric tunneling barrier with a complex-oxide $La_{1-x}Sr_xMnO_3$ magnetic electrode can generate a giant TER effect.

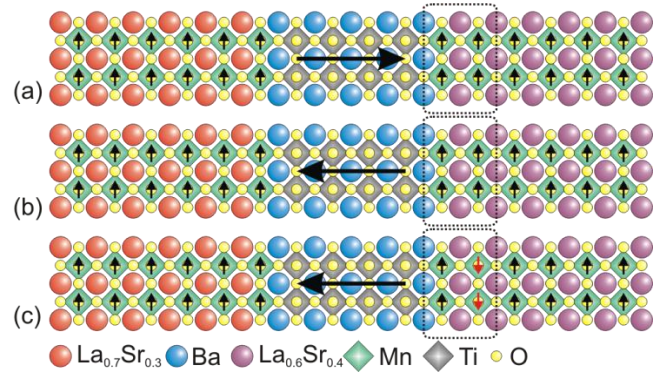

FIG 1. Schematic side view of the layer-by-layer atomic positions and magnetoelectric state of the system. The ferroelectric polarization in the $BaTiO_3$ is indicated by large arrows and the magnetic moment on the Mn sites is indicated by the small arrows. (a) For polarization to the right the Mn magnetic moments are ferromagnetically ordered throughout the right electrode. (b) Configuration with polarization to the left where Mn magnetic moments are assumed to be ferromagnetically ordered everywhere. (c) The ground state magnetic configuration for polarization to the left.

The doped manganite, $La_{1-x}Sr_xMnO_3$, exhibits a rich phase diagram as a function of hole concentration, $x$.[26] Of particular interest for the phenomenon predicted here is the fact that $La_{1-x}Sr_xMnO_3$ exhibits a transition around $x \sim 0.5$ from a ferromagnetic metallic phase to an antiferromagnetic metallic phase.[27] The ferromagnetic phase is half-metallic with electronic density of states only in the majority spin channel.[28] The antiferromagnetic phase has A-type magnetic order consisting of (001) planes of ferromagnetically ordered Mn moments that align antiparallel with neighboring (001) planes. This magnetic structure leads to highly anisotropic "two-dimensional" transport properties: metallic conductivity in the (001) plane and insulating behavior along [001].[29]

In addition to explicit chemical doping, carrier concentration can also be modulated electrostatically, opening the possibility to dramatically alter the electronic properties by ac-tively controlling its electrical boundary conditions,[30] and hence complex oxides such as $La_{1-x}Sr_xMnO_3$ have been the subject of intense investigation in regard to the ferroelectric control of magnetism (see Ref. [24] for a recent review). At the interface of $La_{1-x}Sr_xMnO_3$ with a ferroelectric material, polarization charges on the ferroelectric side of the interface are screened by a build-up of opposite charge on the manganite side, effectively altering the hole concentration near the interface. [25,31] If the manganite has chemical composition $x$ close to a phase boundary then the reversal of polarization can induce a transition locally near the interface. We recently showed that the change in magnetic order around $x \sim 0.5$, discussed above, can be induced by polarization switching at the manganite/ferroelectric interface.[23] Switching polarization leads to a transition from ferromagnetic order to a few unit-cells of antiferromagnetic order, with the A-type ferromagnetic planes parallel to the interface (compare the right interface in Figs. 1a and 1c).

Along with this change in magnetic order comes the transition to the anisotropic metallic phase.[29] This suggests an intriguing possibility to detect this behavior, which constitutes the central result of our Letter: In a FTJ with such a magnetoelectrically active interface in the path of the tunneling current, switching of the ferroelectric barrier is expected to change the "perpendicular metallicity" on the interface of the $La_{1-x}Sr_xMnO_3$ electrode, effectively changing the tunneling barrier thickness, and leading to a giant change in conductance.

We explore this effect using first-principles calculations based on density-functional theory. The junction consists of ~1.9 nm of $BaTiO_3$ as a tunneling barrier with $La_{0.7}Sr_{0.3}MnO_3$ and $La_{0.6}Sr_{0.4}MnO_3$ as the left and right electrodes, respectively.[32] The layers are stacked along the [001] direction of the conventional pseudo-cubic perovskite cell, assuming the typical $AO$-$BO_2$ stacking sequence of perovskite heterostructures (see Fig. 1). We treat the La-Sr substitutional doping using the virtual crystal approximation by considering the $A$-site of the manganite to be occupied by a fictitious atom with non-integer atomic number to reflect the different valence of La and Sr.[33]

Calculations are performed using the plane-wave pseudopotential method implemented in Quantum-ESPRESSO.[34] The supercell used consists of 5 unit cells of $La_{0.7}Sr_{0.3}MnO_3$, 4.5 unit cells of $BaTiO_3$, 5 unit cells of $La_{0.6}Sr_{0.4}MnO_3$ and one additional $MnO_2$ monolayer at the periodic boundary between the $x = 0.3$ and $x = 0.4$ layers. The exchange-correlation effects are treated using the generalized gradient approximation (GGA).[35] All calculations use an energy cutoff of 400 eV for the plane wave expansion. Atomic relaxations of the supercell are performed using a 6×6×1 Monkhorst-Pack grid for k-point sampling and atomic positions are converged until the Hellmann-Feynman forces on each atom became less than 20 meV/Å. The in-plane lattice constant of the supercell is constrained to the calculated (GGA) value for bulk cubic $SrTiO_3$, $a = 3.937$Å, to simulate epitaxial growth on a $SrTiO_3$ substrate, ensuring that the ferroelectric polarization of the $BaTiO_3$ is oriented perpendicular to the plane. Using the Berry phase method we obtain $P = 50\mu C/cm^2$ for the polarization of bulk $BaTiO_3$ in this strain state, which is well within the range achievable experimentally through strain modulation.

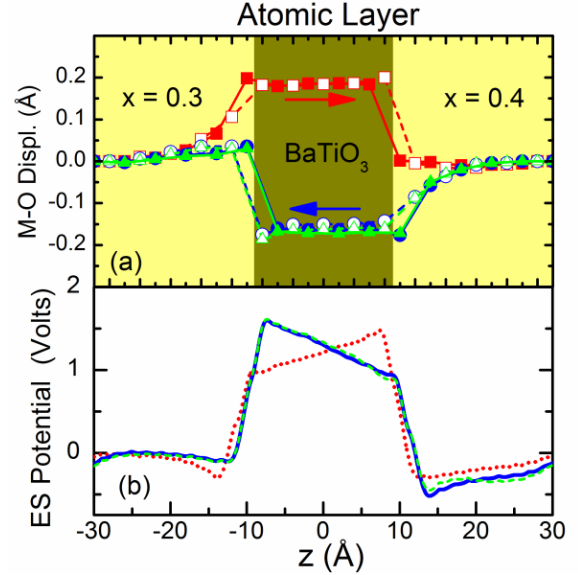

FIG 2. (a) Profile of the relative metal-Oxygen (M-O) displacements in each atomic layer for two polarization states. The solid symbols are $B$-$O_2$ displacements ($B$ = Mn or Ti) and the open symbols are $A$-O displacements (A = $La_{1-x}Sr_x$ or Ba). Squares, circles and triangles correspond to the magnetoelectric states depicted in Fig. 1(a-c), respectively. (b) Macroscopic and planar averaged electrostatic potential profile of the junction. Dotted, solid and dashed curves correspond to the magnetoelectric states depicted in Fig. 1(a-c), respectively.

Ferroelectric displacements in the supercell corresponding to the polarization pointing to right (left) are obtained by fully relaxing the atomic structure under initial conditions of positive (negative) displacements of Ti ions with respect to their neighboring O anions. Subsequent self-consistent calculations of the supercell are performed using a dense, 20×20×1, k-point grid to determine total energies of the different magnetoelectric configurations.

The polarization follows the relative displacement along the $z$-direction of the metal cations (Ba and Ti) with respect to their neighboring oxygen anions in the same formal (001) plane. The square and circle symbols in Fig 2a show these nearly uniform polar displacements throughout the $BaTiO_3$ for polarization to the right and to the left, respectively, assuming the magnetic order in the electrodes is everywhere ferromagnetic (see Fig 1a and b). The triangle symbols are for polarization to the left, but with the ground state magnetic structure depicted in Fig 1c (total energy calculations discussed below establish this magnetic ground state). In addition, polar displacements are also apparent in the $La_{1-x}Sr_xMnO_3$ electrodes resulting from the penetration of the screened electric field due to interface charges.[36]

The lowest energy interfacial magnetic ordering is determined using the same procedure as in Ref [23]. As was pointed out there, only when polarization is pointing away from the interface is the magnetic transition around $x \sim 0.5$ important because this corresponds to additional hole-doping of the inter-face. Since the left electrode has composition far enough away from $x \sim 0.5$, it does not exhibit a phase transition with switching (see note [32]). On the other hand, the right interface is indeed magnetoelectrically active: when polarization is to the right it is ferromagnetically ordered and when polarization is to the left the interface becomes A-type antiferromagnetically ordered in the first 2-3 unit-cells near the interface. While this magnetoelectric state is only metastable (its energy is 72 meV/unit-cell area above the all-ferromagnetic right polarization state in Fig. 1a) its energy is 27 meV/unit-cell area lower than the state depicted in Fig 1b.

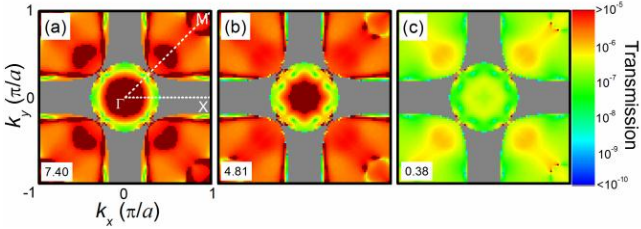

FIG 3. (a-c) The $\mathbf{k}_\parallel$-resolved distribution of the transmission in the 2D Brillouin zone for different magnetoelectric states depicted in Fig 1(a-c), respectively. The integrated total conductance G [see eq (1)] is given in the lower-left corner of each plot (in units of $10^{-6} e^2/h$). Note that the color scale is logarithmic and the gray areas correspond to zero transmission.

The conductance per unit cell area is given by the Landauer-Büttiker formula

$$G_\sigma = \frac{e^2}{h} \sum_{\mathbf{k}_\parallel} T_\sigma(\mathbf{k}_\parallel) , \qquad (1)$$

where $T_\sigma(\mathbf{k}_\parallel)$ is the transmission probability of an electron at the Fermi energy with spin σ and Bloch wave vector $\mathbf{k}_\parallel = (k_x,k_y)$. The tunneling transmission is calculated using a general scattering formalism implemented in the Quantum-ESPRESSO package.[34] The structures depicted in Fig. 1 are considered as a central scattering region ideally attached on both sides to semi-infinite $La_{1-x}Sr_xMnO_3$ leads. Matching the wave functions of the scattering region at the interfaces to the propagating states in semi-infinite $La_{1-x}Sr_xMnO_3$ electrodes yields transmission coefficients. The two-dimensional Brillouin zone is sampled using a uniform 100×100 $\mathbf{k}_\parallel$ mesh. In the present case only the majority-spin channel has non-zero transmission because of the half-metallic behavior of both $La_{1-x}Sr_xMnO_3$ electrodes. The resulting transmission distributions, $T_\sigma(\mathbf{k}_\parallel)$, for the three magnetoelectric states depicted in Fig. 1 are plotted in Fig. 3.

All three figures consist of grey areas corresponding to the regions of the Fermi surfaces of either metal electrode where there are no available conducting states. Comparing Fig. 3a and b for the two states depicted in Fig. 1a and b, respectively, we find a noticeable difference in the transmission distribution. To quantify the change in transport properties with polarization switching, we define a quantity known as the TER ratio,

$$TER = \frac{G_R - G_L}{G_L} \times 100\% , \qquad (2)$$

in terms of the conductance G of the right (R) and left (L) polarization states. We find a modest ratio of 54%. The origin of this change is the shift of the average electrostatic potential in the $BaTiO_3$ barrier with polarization reversal, as can be seen in Fig. 2b, altering the height of the effective barrier through which electrons must tunnel.[7]

More interesting, though, is the comparison of the transmission in Fig. 3a and in Fig. 3c. The transmission plotted in Fig. 3c includes the antiferromagnetic insulating layer at the interface (see Fig 1c). Correspondingly the transmission throughout the entire Brillouin zone is significantly depressed compared to the case where polarization points to the right. The total conductance changes by more than an order of magnitude and in this case TER ~1800%.

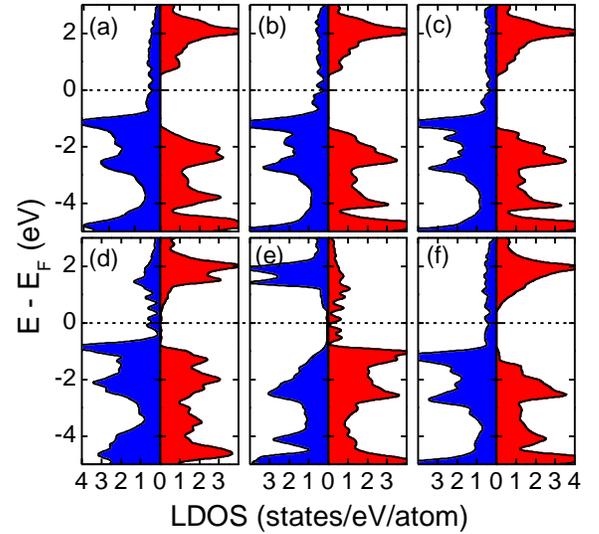

FIG 4. (a)-(c) Local density of states (LDOS) of the first, second and third unit-cells of $La_{0.6}Sr_{0.4}MnO_3$, respectively, near the interface for polarization pointing to the right. The left (right) panels are for spins parallel (antiparallel) to the bulk magnetization. (d)-(f) The same as in (a)-(c) except for ferroelectric polarization to the left.

This large change in conductance cannot be accounted for by a change in electrostatics of the barrier. The electrostatic potential is not significantly different from that of the case with a ferromagnetic interface (compare solid and dashed curves in Fig. 2b), and therefore the contribution to the effective tunneling barrier from the $BaTiO_3$ does not change appreciably. To understand the effect the change in magnetic order has on the transmission we compare the spin-polarized local density of states (LDOS) on the interfacial layers of the right electrode for the two polarization states, plotted in Fig. 4. For polarization to the right, the LDOS at the Fermi level is non-zero only in the majority-spin channel due to the half-metallicity. After polarization reversal, however, the antiferromagnetic transition sets in. This change in magnetic order reverses the moment of the second Mn layer, (compare Fig. 1a and c) which interchanges the role of "up" and "down" spin-states in the second unit-cell (compare Fig. 4b and e). Because of the large spin-polarization of the Mn sites the LDOS for majority-spin electrons at the Fermi level is very small in the second unit-cell. Therefore the first and second unit-cells at the interface can be considered atomic scale "electrodes" separated by a monolayer-thin spacer layer, i.e. an atomic-scale spin-valve. Since the spin-polarization is nearly 100% this "spin-valve" exhibits a huge difference in conductance between parallel and anti-parallel configurations.

The predicted effect can be established experimentally by fabricating a series of FTJs and varying the nominal Sr concentration, $x$, of one of the electrodes, with the other electrode magnetoelectrically inert.[32] Below the critical $x$ the TER will be modest, while above the critical $x$ it will increase dramatically due to the onset of the spin-valve effect at the interface. Thus, we hope that our theoretical predictions will stimulate experimental studies of the predicted giant TER effect in FTJs with $La_{1-x}Sr_xMnO_3$ electrodes.

This work was supported by the Nebraska MRSEC (NSF Grant No. DMR-0820521), the Nanoelectronics Research Initiative, NSF-EPSCoR (Grant No. EPS-1010674) and the Nebraska Research Initiative. Computations were performed at the Holland Computing Center of the University of Nebraska.

[*] e-mail: jdburton1@gmail.com
[**] e-mail: tsymbal@unl.edu